\documentclass[11pt]{article}
\usepackage[utf8]{inputenc}
\usepackage{xspace}
\usepackage{setspace}
\usepackage{fullpage}
\usepackage{graphicx}
\graphicspath{ {./figures/} }
\usepackage{subcaption}
\usepackage{amsmath,amsfonts,amssymb}
\usepackage[colorlinks,citecolor=blue,bookmarks=true,linktocpage]{hyperref}
\usepackage{xcolor}
\usepackage{fullpage}
\usepackage{amsthm}
\usepackage{algorithm}
\usepackage{algpseudocode}
\usepackage{bbm}
\usepackage{mathtools}
\usepackage{float}
\usepackage{indentfirst}
\usepackage{tikz}
\usepackage{accents}
\usetikzlibrary{arrows.meta}
\usetikzlibrary{shapes.misc}
\tikzset{cross/.style={cross out, draw=black, minimum size=2*(#1-\pgflinewidth), inner sep=0pt, outer sep=0pt},
cross/.default={1pt}}
\allowdisplaybreaks
\newtheorem{theorem}{Theorem}[section]

\newtheorem{lemma}{Lemma}[section]

\newtheorem*{prop*}{Proposition}

\theoremstyle{definition}
\newtheorem{definition}{Definition}[section]

\theoremstyle{remark}
\newtheorem{remark}{Remark}
\numberwithin{equation}{section}
\usepackage[capitalize]{cleveref}

\newcommand{\opt}{\operatorname{OPT}}
\newcommand{\cost}{\operatorname{cost}}

\newcommand{\dist}{\operatorname{dist}}
\newcommand{\supp}{\operatorname{supp}}
\newcommand{\etal}{\emph{et al.\@}}
\usepackage[backend=biber,style=alphabetic,maxbibnames=99,maxalphanames=99,url=false]{biblatex}
\providecommand{\email}[1]{\href{mailto:#1}{\nolinkurl{#1}\xspace}}

\title{With a Little Help From My Friends: Exploiting Probability Distribution Advice in Algorithm Design}

\author{
Cl\'ement L. Canonne\thanks{The University of Sydney.
Email: \email{clement.canonne@sydney.edu.au}.}
\and
Kenny Chen\thanks{The University of Sydney.
Email: \email{kche5493@uni.sydney.edu.au}.}
\and 
Juli\'an Mestre\thanks{The University of Sydney. Email: \email{julian.mestre@sydney.edu.au}.}
}

\begin{document}

\maketitle

\begin{abstract}
We study online algorithms with predictions using \emph{distributional advice}, a type of prediction that arises when leveraging expert knowledge or historical data. To demonstrate the usefulness and versatility of this framework, we focus on the fundamental problem of \emph{online metric matching}, considering both the fractional and integral variants. Our main positive result is, for the former, an algorithm achieving the optimal cost under perfect advice, while smoothly defaulting to competitive ratios comparable to advice-free algorithms as the prediction's quality degrades. For the integral matching, we are able to provide an algorithm with essentially the same guarantees, up to an additive sublinear term. We conclude by discussing how our algorithmic framework can be extended to other online optimization problems.
\end{abstract}

\section{Introduction}
\emph{Algorithms with predictions} (or algorithms with advice) is a form of beyond-worst-case analysis first introduced by Lykouris and Vassilvitskii in 2018 \cite{DBLP:conf/icml/LykourisV18}. Compared to classical analysis, the framework of algorithms with advice also provides the algorithm with a prediction, crucially not guaranteed to be correct, as additional input. The aim is then to design algorithms which can utilize these predictions to achieve results which are better than classical algorithms when the predictions are ``good,'' while not performing much worse than classical algorithms otherwise. A natural class of algorithms which have been studied with predictions are \emph{online algorithms}, where uncertainty about the future is a large part of what makes problems difficult (see, e.g, \cite{DBLP:journals/talg/AntoniadisCEPS23,DBLP:journals/disopt/AntoniadisGKK23,DBLP:conf/innovations/LindermayrMS22}).

Most of the existing literature has been focused on predictions which are tailored specifically to the problem and non-probabilistic, such as the maximum value of all the ``secretaries'' in the online secretary problem \cite{DBLP:journals/disopt/AntoniadisGKK23}, the color of a vertex in the online graph coloring problem \cite{DBLP:conf/iscopt/AntoniadisBM24} or the number of days a person wishes to ski for in the (multi-shop) ski rental problem~\cite{DBLP:conf/nips/WangLW20}. In practice, however, it can be highly impractical or unrealistic to obtain this type of prediction. For example, knowing the maximum value of all secretaries would involve screening all the secretaries beforehand, or obtaining a prediction for vertex coloring would require knowing a coloring in advance. For many optimization problems, it would be a tall order to ask for a predictive model that not only can predict the input ahead of time but can also compute the optimal solution. This motivates the question of whether we can have algorithms that can exploit more attainable forms of predictions as input.

One alternative type of prediction is a \emph{distributional prediction}, where the prediction is a probability distribution meant to convey some information about the input. This type of prediction has multiple advantages. For one, it is much less burdensome to obtain. For example, consider the problem ride-hailing companies face when pairing drivers with hailers: Though they may not know exactly where hails will come from, they have access to data which they can use to estimate distributions for where hails will come from at specific times, like peak hours. Furthermore, this prediction model aligns well with machine learning models and neural networks, which have recently seen tremendous growth, since the natural output of such models is a probability distribution. Finally, the setting is very general. It can be easy to imagine a company using prior hiring cycles to obtain a baseline distribution for the graduates they are about to interview, or a company using their current data to estimate where the best places for a possible expansion to capture as many new customers as possible should be. In all of these examples, it is easy to see how the supplied prediction distribution can differ from the actual input distribution, meaning it is of real interest to design algorithms which can utilize the predictions to obtain performance better than the worst case if they are accurate, but still achieve some baseline performance guarantee if they are not.

 This type of prediction was first studied by Diakonikolas~\etal~\cite{DBLP:conf/icml/DiakonikolasKTV21} for the online ski rental and prophet inequality problems. Instead of knowledge of the underlying distribution though, they allowed their algorithm to sample from an unknown prediction distribution. The idea was further pursued in Angelopolous~\etal~\cite{ijcai2024p404} and Dinitz~\etal~\cite{DBLP:conf/nips/DinitzILMNV24}, where algorithms for contract scheduling and binary searching respectively were equipped with full knowledge of the distribution from which the actual input would be drawn, rather than just being allowed to query an unknown distribution. To the best of our knowledge, these were the first instances where the prediction being a distribution over possible inputs was explored.

In this paper, we extend this idea to the \emph{online} setting, where the algorithm is given a probability distribution as a prediction. This prediction can either be a distribution which purports to be as similar to the probability distribution online arrivals will be drawn from, if in the stochastic setting, or it can aim to be similar to the empirical distribution induced by the online arrivals, if otherwise.

The main task we study is the \emph{online metric matching problem}. Here, there are $N$ servers given offline on a metric space, after which $N$ online requests arrive. As each request arrives, the algorithm must irrevocably match the request with servers, with the goal being to minimize the cost of the overall matching once all requests arrive. The prediction we use is a probability distribution over the metric space, purported to model the likely locations of the $N$ incoming requests. We first introduce and study the fractional variant of this problem (where a request can be fractionally assigned to different servers), before extending the idea to the more common integral version (where a request must be assigned to a single server). We then show how the same algorithmic ideas can be extended to a broader class of metric problems.
\subsection{Our Results} 
In view of describing our results, we first briefly recall the definition of $1$-Wasserstein distance (also known as \emph{Earthmover distance}; see~\cref{sect:preliminaries}), and motivate it as our measure of error. At a high level, the 1-Wasserstein distance, henceforth Wasserstein distance for simplicity, represents the minimum cost needed to turn one probability distribution into another, with the cost measured by the probability mass moved across the metric space: moving a unit of probability mass from point $x$ to point $y$ has cost given by the distance between $x$ and $y$. In this sense, the offline minimum metric matching problem can be reformulated as computing the Wasserstein distance between the empirical distributions induced by the request and server sets, up to scaling factors. This distance is thus a natural choice for our error measure, as it encapsulates the task's objective, and incorporates the salient characteristics of the metric space. We expand on this more in \cref{sect:notion_of_advice}.

We show in \cref{sect:hardness_fractional_randomized} that under adversarial arrivals, the fractional online metric problem is not easier than the well-studied integral variant:
\begin{theorem}[Informal, see~\cref{thm:fractional_matching_hardness}]\label{inf:thm:fractional_matching_hardness}
Every randomized algorithm for online fractional metric matching must have competitive ratio $\Omega(\log N)$ in the adversarial arrival setting. This lower bound matches the hardness results for online \emph{integral} metric matching.
\end{theorem}
Given that our next result is concerned with the online fractional matching task, but using an algorithm for integral matching as a blackbox, the above ensures that we are comparing apples to apples, and not implicitly paying some suboptimal approximation ratio in the move from integral to fractional matching.  
In more detail, the algorithm with distributional predictions for online fractional metric matching we design has the following guarantees:
\begin{theorem}[Informal, see~\cref{thm:fractional_matching_guarantees}]
    \label{thm:fractional_matching_guarantees:informal}
 There is an algorithm for online fractional metric matching in the adversarial arrival setting with competitive ratio $1 + \frac{\beta\cdot\eta+o(1)}{\opt}$, where $\eta$ is the Wasserstein distance between the advice and true distributions, $\beta$ is the best possible competitive ratio for the online \emph{integral} metric matching, and $\opt$ is the cost of the optimal matching between the servers and online requests. 
 In particular, the algorithm asymptotically achieves competitive ratio 1 when the advice is perfect.
\end{theorem}
Notably, this algorithm can use any baseline online algorithm as a black box, and is not dependent on the behavior of any specific algorithm. Furthermore, this also means that the performance of this algorithm will improve along with improvements in the adviceless online algorithms.

Unlike previous works on online matching where the consistency is at least $1+\Omega(1)$, ours is asymptotically optimal. Moreover, we show in~\cref{app:hardness_result} (\cref{thm:integer_matching_advice_hardness_result}) that the additional (vanishing) $o(1)$ term is necessary.

\medskip

We then show how the algorithmic ideas underlying~\cref{thm:fractional_matching_guarantees:informal} can be extended to the online integral metric matching problem:
\begin{theorem}[Informal, see~\cref{thm:integer_metric_matching_algorithm}]
 There is an algorithm for online integral metric matching in $\mathbb{R}^d$ in the adversarial arrival setting with competitive ratio {$1+\frac{\beta\cdot\eta+\widetilde{O}(N^{1/2})}{\opt}$}, where $\eta$ is the Wasserstein distance between the advice and true distributions, $\beta$ is the best possible competitive ratio for the online \emph{integral} metric matching, and $\opt$ is the cost of the optimal matching between the servers and online requests.
\end{theorem}
The main difference between the guarantees obtained in the fractional and integral cases is the presence, in the latter, of an additional $\widetilde{O}(N^{1/2})$ term in the competitive ratio. While sublinear in $N$, this term does not vanish as the number of requests grows: unfortunately, it seems inherent to our approach, and arises from the first step of our algorithm, which consists of the conversion from a (possibly continuous) advice distribution over an arbitrary support to a finite multiset. More specifically, it quantifies the convergence rate between the empirical distribution over this multiset and the original advice distribution, in Wasserstein distance. We also describe how to remove the assumption of the Euclidean metric space, at the cost of polylogarithmic factors.

\subsection{Our Techniques}
The high-level idea of our matching algorithm is as follows. Given a distribution provided as advice, our algorithm will produce a set of $N$ points from it, and use those points to simulate the online arrivals it is about to receive. After obtaining these advice points, it can compute the optimal matching between the samples and the servers offline. Then, as the actual requests arrive, the algorithm can instead perform an online matching between these true requests and these advice points, the intuition being that if the advice is indeed a good representation of the requests, the algorithm can circumvent the difficulty of the problem, which stems from not knowing the future online arrivals. An example of how the algorithm behaves on an example on the line (i.e., in $\mathbb{R}$) is given in \cref{fig:samplingexample}.
\begin{figure}[H]
    \centering
    \begin{minipage}{0.32\textwidth}
    \centering
        \begin{tikzpicture}[>=Stealth, scale=0.7, transform shape]
            \begin{scope}[xshift=-4cm, local bounding box=dist1]
            \draw[<->, thick] (-3.5,0) -- (3.5,0);
            \foreach \x/\label in {-2.5/s_1, -1/s_2, 0.5/s_3, 1.5/s_4, 2.8/s_5}
                \filldraw[red] (\x,0) circle (1.6pt) node[below, yshift=-3pt, black, font=\small] {$\label$};
            \draw[blue, thick, smooth, domain=-3.5:3.5, samples=100] plot (\x, {2*exp(-\x*\x/2)}); %
            \filldraw[white] (0,-1) circle (0pt);
            \draw[white] (0,-3) circle (0pt);
            \end{scope}
        \end{tikzpicture}
    \end{minipage}
    \begin{minipage}{0.32\textwidth}
        \centering
        \begin{tikzpicture}[>=Stealth, scale=0.7, transform shape]
            \begin{scope}[xshift=0cm, local bounding box=dist2]
            \draw[<->, thick] (-3.5,0) -- (3.5,0);
            \foreach \x in {-2.5, -1, 0.5, 1.5, 2.8}
                \filldraw[red] (\x,0) circle (1.6pt); %
            \foreach \x/\label in {-0.921/p_1, -0.542/p_2, 0.314/p_3, 0.763/p_4, 1.678/p_5}
                \filldraw[black] (\x,-1) circle (1.6pt) node[below, yshift=-3pt, font=\small] {$\label$};; %
            \draw[blue, thick, smooth, domain=-3.5:3.5, samples=100] plot (\x, {2*exp(-\x*\x/2)});
            \draw (-0.921,-1) -- (-2.5,0);
            \draw (-0.542,-1) -- (-1,0);
            \draw (0.314,-1) -- (0.5,0);
            \draw (0.763,-1) -- (1.5,0);
            \draw (1.678,-1) -- (2.8,0);
            \draw[white] (0,-3) circle (0pt);
            \end{scope}
        \end{tikzpicture}
    \end{minipage}
    \begin{minipage}{0.32\textwidth}
        \centering
        \begin{tikzpicture}[>=Stealth, scale=0.7, transform shape]
            \begin{scope}[xshift=4cm, local bounding box=dist3]
            \draw[<->, thick] (-3.5,0) -- (3.5,0);
            \foreach \x in {-2.5, -1, 0.5, 1.5, 2.8}
                \filldraw[red] (\x,0) circle (1.6pt); %
            \draw[blue, thick, smooth, domain=-3.5:3.5, samples=100] plot (\x, {2*exp(-\x*\x/2)});
            \foreach \x in {0.763, -0.921, 0.314, -0.542, 1.678}
                \filldraw[black] (\x,-1) circle (1.6pt); %
            \draw[blue, thick, smooth, domain=-3.5:3.5, samples=100] plot (\x, {2*exp(-\x*\x/2)});
            \draw (-0.921,-1) -- (-2.5,0);
            \draw (-0.542,-1) -- (-1,0);
            \draw (0.314,-1) -- (0.5,0);
            \draw (0.763,-1) -- (1.5,0);
            \draw (1.678,-1) -- (2.8,0);
            \draw[black] (-1,-2) circle (1.6pt) node[anchor=north]{$r_{1}$};
            \draw (-1,-2) -- (-0.921,-1);
            \draw[white] (0,-3) circle (0pt);
            \end{scope}
        \end{tikzpicture}
    \end{minipage}\vspace{-1em}
    \caption{\small{}Walkthrough of how the algorithm works on an example of online metric matching on the line. For visual clarity, the sequence of points are presented with differing $y$ axis values, but they all occur on the line. Here, the \textcolor{red}{red} points on the line are the offline servers, labeled $s_{i}$, and the \textcolor{blue}{blue} normal curve is the advice distribution. The algorithm obtains $N$ advice points from the advice distribution (in this case $N=5$, labeled $p_1,\dots, p_5$) which are represented by the black points in the second figure. It then pretends these points are the offline requests, and computes the optimal offline matching between the advice points and the servers. Then, as online requests come in, represented by the open circles in the third figure, it uses a black box algorithm to determine which of the advice points to match to, and overall matches it with the corresponding server accordingly. The first online request is denoted $r_{1}$, which is matched to $p_{1}$. Hence, our overall algorithm with advice matches $r_{1}$ to $s_{1}$.}
    \label{fig:samplingexample}
\end{figure}
We show that the Wasserstein distance, which as previously discussed we use as our measure of error for the advice, can be related to the expected cost of an optimal matching between the real requests and the distributional prediction, from which we are able to compare our prediction distribution against the empirical distribution induced by the sequence of requests using the Wasserstein distance. There is an intricacy here, where we need to sample enough points so that the Wasserstein distance between the empirical distribution induced by the samples and the advice distribution converges sufficiently fast. We show that this is achievable for the fractional online metric matching problem, but unfortunately not for the integral metric matching problem, except for special cases~--~explaining the necessity of the aforementioned additive term. These results are first shown for metric spaces in $\mathbb{R}^{d}$, then for general metric spaces by embedding them into $\mathbb{R}^{d}$ with sufficiently small distortion.

\subsection{Other Related Work}
The online integral metric matching problem has been extensively studied in a range of contexts. Two of the main variants are the \emph{adversarial} arrival and \emph{random} arrival model, the latter introduced in~\cite{GoelM08}, which differ in the assumptions made about the sequence of requests. In the adversarial case, an adversary with full knowledge of the algorithm controls both the set of requests and the order of their arrival, meaning that they can adaptively choose the next requests in response to previous decisions made by the algorithm. In contrast, for the random arrival model, although the adversary has full knowledge of the algorithm, it must fix a set of requests prior to the execution of the algorithm, after which the order of their arrival is random. For adversarial arrivals, Khuller~\etal~\cite{DBLP:journals/tcs/KhullerMV94} obtained a $(2n-1)$-competitive algorithm on general metric spaces, a competitive ratio they showed to be tight for deterministic algorithms in this arrival model. When allowing randomization, this can improved to an expected $O(\log^{2}n)$-competitive ratio \cite{DBLP:journals/algorithmica/BansalBGN14}, complemented by an $\Omega(\log N)$ lower bound for randomized algorithms in the adversarial arrivals model \cite{DBLP:conf/soda/MeyersonNP06}.  For the random arrivals model, Raghvendra  presented a deterministic algorithm \cite{DBLP:conf/approx/Raghvendra16}, which they subsequently showed to be $O(\log N)$-competitive \cite{DBLP:conf/compgeom/Raghvendra18}. In the randomized setting, Gupta~\etal~\cite{DBLP:conf/icalp/GuptaL12} provide an $O(\log n)$ expected competitive algorithm for doubling metrics. These results are summarized in \cref{tab:compratios}.

\begin{table}[H]
    \centering 
    \scalebox{1}{
        \begin{tabular}{c||c|c}
             ~ & Adversarial Arrival & Random Arrival \\
             \hline 
             Deterministic Algorithm & $O(N)$ \cite{DBLP:journals/tcs/KhullerMV94} & $O(\log N)$ \cite{DBLP:conf/compgeom/Raghvendra18}\\
             \hline
             Randomized Algorithm & $O(\log^{2} N)$ \cite{DBLP:journals/algorithmica/BansalBGN14} & $O(\log N)$ (doubling metrics) \cite{DBLP:conf/icalp/GuptaL12} \\
             \hline
        \end{tabular}
    }
    \caption{Competitive ratios $\beta$ for deterministic and randomized algorithms under the adversarial and random arrival settings for general metrics.}
    \label{tab:compratios}
\end{table}
It is also worth noting that the above are for general metrics, and better results can be obtained on specific metrics. For example, Raghvendra~\cite{DBLP:conf/compgeom/Raghvendra18} showed the same $O(\log N)$-competitive algorithm for random arrivals also achieved the same competitive ratio for adversarial arrivals specifically on the line metric.

Turning to online metric matching with advice, Antoniadis~\etal~\cite{DBLP:journals/talg/AntoniadisCEPS23} give a deterministic algorithm that achieves a competitive ratio of $\min\{2n-1,9+\frac{8e\eta}{\opt}\}$. The prediction this algorithm uses is, for every online request arrival, an indicator of which server it should match to. For any given request $i$, let $\eta_{i}$ be the absolute difference in costs between the matching edge induced by the prediction, and that induced by an optimal solution. For example, for request $r_{i}$, if the prediction indicates this request should be matched to $s_{p}$, but in an optimal solution it is matched to $s_{o}$, then $\eta_{i}=|d(r_{i},s_{p})-d(r_{i},s_{o})|$. Then, $\eta$ is defined as $\eta = \max_{i} \eta_i$, that is, the maximum difference in cost, compared to an optimal solution, contributed by a single request. In contrast, our algorithm is a randomized algorithm which achieves slightly better consistency, at the cost of a $\beta$ factor on (our) error term $\eta$, instead of constant (based on \cref{tab:compratios}, $\beta$ will mainly be $O(\log N)$ for our purposes). However, comparing a randomized algorithm with a deterministic algorithm which relies on a very strong and incomparable type of advice is not the most meaningful benchmark. The most interesting and comparable variant of online metric matching for us is the stochastic online metric matching variant studied by Gupta~\etal~\cite{DBLP:journals/corr/abs-1904-09284}. Here, the online requests are drawn from a known distribution, and the authors provide a $O((\log\log\log N)^{2})$-competitive algorithm for general metrics, shown to be $9$-competitive for line and tree metrics specifically.

Concurrently and independently from us, Yang~\etal~\cite{yang2025onlinemetricmatchingworst} give a switching algorithm for online integral metric matching under the random arrivals model with predictions that uses similar ideas to ours but with one key difference: They assume the prediction is a set of points representing a putative set of request (up to ordering). In this model, they give an $O(1)$-competitive under perfect advice and $O(N)$-competitive otherwise under random arrivals, building on the switching algorithm of Antoniadis~\etal~\cite{DBLP:journals/talg/AntoniadisCEPS23}. These results can be seen as a special case of ours, by converting their multiset prediction into the uniform distribution induced by the multiset, in which case our algorithm obtains similar guarantees. We further note that algorithm can be made $O(N)$-competitive under imperfect advice using the same switching approach as theirs, though this inherently comes at the cost of trading the $1$-competitive guarantee under perfect advice for an $O(1)$-competitive guarantee.

\section{Preliminaries}\label{sect:preliminaries}
We use standard asymptotic big-Oh notation throughout, as well as the (slightly) less standard $\tilde{O}$ notation, defined by $\tilde{O}(f(N))=O\left(f(N)\log^{k}f(N)\right)$, for some integer $k$. Let $I$ be an instance of a problem, with optimal solution $\opt$ on the instance. Then, we measure the performance of an algorithm by its competitive ratio:
\begin{equation*}
    c=\frac{\cost(M)}{\opt}.
\end{equation*}
For randomized algorithms, we analogously consider the \emph{expected} competitive ratio, replacing in the above the numerator with $\mathbb{E}[\text{cost}(M)]$. We say that an algorithm is \emph{$c$-competitive} if it attains a competitive ratio of $c$. From now on, we will use the terms \emph{advice} and \emph{prediction} interchangeably.

\paragraph{The Metric Matching Problem.}
The \emph{offline integral metric matching} problem is defined as follows. On a metric space $(X,\dist)$, the algorithm is given a set of requests, $R$, and a set of servers $S$, both of cardinality $N$. They must find a configuration that pairs each request $r$ to some server $s$ while minimizing the overall cost, which is given by the sum of the distances between the assigned pairs, denoted $\dist(r,s)$. Letting $x_{r,s}$ be the indicator variable for whether the algorithm pairs request $r$ with server $s$, this can be formulated as the following integer linear program:
\begin{align*}
    \min&\sum_{r\in R,s\in S}\dist(r,s)x_{r,s},\\
    \text{subject to:}&\\
    &\sum_{s\in S}x_{r,s} \geq 1,~\forall r\in R\\
    &\sum_{r\in R}x_{r,s} \leq 1,~\forall s\in S\\
    &x_{r,s} \in \{0, 1\},~\forall r\in R, s\in S.
\end{align*}
We denote the cost of the optimal solution to the above integer program on request set $R$ and server set $S$ as $\opt(R,S)$, or sometimes $\opt_{I}(R,S)$, to emphasize the integral matching.
\\\indent We can also relax the integrality constraints to the above (the third constraint) to allow any real value $0 \leq x_{r, s} \leq 1$, which results in the \emph{offline fractional metric matching} problem. In other words, we allow the algorithm to fractionally pair requests with corresponding servers. We denote the cost of the optimal matching here as $\opt(R,S)$, or sometimes $\opt_{F}(R,S)$, when we wish to emphasize the fractional matching. In this setting, we let the \emph{cost} of the edge be the distance between the request and the server (i.e., $\dist(r,s)$) and the \emph{weight} of the edge be the fractional portion allocated to the edge (i.e., $x_{r,s}$). For any server $s$, we refer to the sum of all the weights of the fractional edges incident on it as the \emph{capacity} of $s$. The following is a well-known result concerning the bipartite matching polytope:
\begin{theorem}[Birkhoff–von-Neumann]\label{thm:bipartite_polytope}
    For any request set $R$ and server set $S$:
    $
        \opt_{I}(R,S)=\opt_{F}(R,S).
   $
\end{theorem}
In other words, the cost of the optimal integral matching and the fractional matching are the same. When it is clear from the context, we sometimes also use $\opt(R,S)$, $\opt_I(R,S)$ and $\opt_F(R,S)$ to denote the actual matching as well.

For the \emph{online} variation of these problems, the algorithm is given the server set $S$ upfront, but the request set $R$ arrives in an online manner. Upon the arrival of each request, the algorithm must irrevocably pair the request to servers in the server set. In the integral variant, the request must be paired to exactly one server, which cannot be paired again. In the fractional variant, the request may be fractionally paired to multiple servers, as long as the sum of fractional edges incident on the request totals one, and no server has sum of fractional edges more than one.
\paragraph{Consistency and Robustness.} The objective of algorithms with advice is to design an algorithm which obtains close to the optimal solution if the advice given is good, and not much worse than the best online algorithm without advice. The usual way to analyze such an algorithm is using the consistency/robustness framework, with some appropriate notion of error, which we will denote with the letter $\eta$. The specific formulation of $\eta$ differs depending on the problem and the advice model, but the aim is for it to capture how different the given prediction is from the true set of online inputs. The \emph{consistency} of an algorithm is the competitive ratio which it obtains when given a perfect prediction. The \emph{robustness} of an algorithm is the competitive ratio (possibly dependent on $\eta$, and ideally degrading smoothly with it) it obtains when given an otherwise bad prediction.
\paragraph{Distributional Notions of Distance.} When considering how far the advice is from being good, the measure of distance we will use for is the \emph{1-Wasserstein distance} (also known as the \emph{earthmover distance}), or simply in this paper \emph{Wasserstein} distance.
\begin{definition}
    Let $X$ and $Y$ be random variables on a metric space $M$, with probability marginals $\mu$ and $\rho$ respectively. Then the \emph{1-Wasserstein distance} is given by
    \begin{equation*}
        W_{1}(\mu, \rho)=\inf\left\{\int_{M\times M}|x-y|\xi(dx,dy):~\xi\in\mathcal{H}(\mu,\rho)\right\},
    \end{equation*}
    where the infimum is over the set of all couplings of random variables $X$ and $Y$ i.e., the set of all joint probability measures with marginals $\mu$ and $\rho$, denoted $\mathcal{H}$.
\end{definition}
We will also later make use of the following result on the convergence between a distribution and the empirical distribution induced by $N$ samples from it:
\begin{lemma}[Fournier and Guillin \cite{fournier2013rateconvergencewassersteindistance}]\label{lem:wasserstein_convergence_rate}
    Let $\rho$ be a probability measure on a space $\mathcal{M}\subseteq \mathbb{R}^{d}$ with finite q-th moment $M_{q}(\rho)$ for some $q\geq 1$, and let $\widehat{\rho}_{N}$ be its empirical measure,
    \begin{equation*}
        \widehat{\rho}_{N}=\frac{1}{N}\sum_{i=1}^{N}\delta_{x_{i}},
    \end{equation*}
    for samples $X_{N}=(x_{1},...,x_{N})\sim \rho^{N}$. Then,
    \begin{equation*}
        \mathbb{E}[W_{1}(\rho,\widehat{\rho}_{N})]=\begin{cases}
            O\left(N^{-1/2}\right),&d=1\\
            O\left(N^{-1/2}\cdot\log N\right),&d=2\\
            O\left(N^{-1/d}\right),&d\geq3.
        \end{cases}
    \end{equation*}
\end{lemma}

\section{Online Fractional Metric Matching}
We now move onto our algorithm taking distributional advice for fractional matching. We first define the form of advice, before describing and analyzing our algorithm for the fractional variant, and establishing a hardness result. 
\subsection{Notion of Advice}\label{sect:notion_of_advice}
The algorithm will be given a prediction $\mathcal{A}$, which aims to inform the algorithm of the incoming set of requests. Let $R$ be the set of requests; with a slight abuse in notation we also use $R$ to denote the empirical distribution induced by the set of requests. Then, we define our error as the Wasserstein distance between the prediction $\mathcal{A}$ and the empirical distribution induced by the set of requests, $\eta=W_{1}(\mathcal{A},R)$.
To see how the Wasserstein distance really is computing the optimal matching up to a scaling factor, first consider the simplified case where $\mathcal{A}$ is a uniform distribution over $N$ points. Then, by definition of the Wasserstein distance, $\eta$ is defined as:
\begin{equation*}
    \eta=W_1(\mathcal{A},R)=\inf_{\pi}\left(\frac{1}{N}\sum_{i=1}^{N}\left|\mathcal{A}_i-R_{\pi(i)}\right|\right),
\end{equation*}
where $\mathcal{A}_i$ and $R_i$ represent the elements in the support of $\mathcal{A}$, and the request set respectively, and $\pi$ is a permutation on $N$ elements. In other words, this is the cost of the optimal matching between the elements in the support of $\mathcal{A}$ and the request set, up to a scaling factor of $N$. This same intuition can be generalized for distributions $\mathcal{A}$ with arbitrary support size, in which case the Wasserstein distance measures the optimal fractional matching between the distribution $\mathcal{A}$ and the request set $R$, up to the same scaling factor of $N$.
\subsection{Algorithm for Fractional Matching}
We now present our algorithm for online fractional matching. We first assume the problem occurs on the metric space $\mathbb{R}^d$; at the end of this section we explain how this assumption can be relaxed. For this problem, the algorithm is given offline the server set $S$, an advice distribution $\mathcal{A}$, and will receive online the request set $R$. The presented algorithm has the following guarantees.
\begin{theorem}\label{thm:fractional_matching_guarantees}
    There is an algorithm for online fractional matching against an adversary that is an asymptotically $1+o(1)$-consistent and $\left(1+\frac{\widetilde{O}(d^2N)\cdot\eta+o(1)}{\opt}\right)$-robust algorithm in expectation, where $d$ is the dimension of the metric space. Moreover, $d^2$ can be replaced with $\min\{d^2,\log^2 N\}$.
\end{theorem}
We show the necessity of an $o(1)$ term in \cref{app:hardness_result}. Before processing any online requests, the algorithm does the following. First, it samples $N^{d+1}$ points from the advice distribution $\mathcal{A}$. Denote the empirical distribution induced by these points and the multiset of these sampled points as $\widetilde{D}$. Next, for every server in $S$, it creates $N^{d}$ copies of it, forming the multiset $\widetilde{S}$, with cardinality also $N^{d+1}$. Next, it computes the optimal matching between the set $\widetilde{D}$ and $\widetilde{S}$, with cost $\opt_{I}(\widetilde{D},\widetilde{S})$. After this, it initializes some randomized online integral metric matching algorithm (call it $\mathcal{B}$), with competitive ratio $\beta$ and lets the server set be the sample of points $\widetilde{D}$. The algorithm then starts to process the online requests. As each online request $r$ arrives, it feeds $N^{d}$ online arrivals (``copies'' of this point) into $\mathcal{B}$. Hence, this matches the request $r$ to $N^d$ points in the set $\widetilde{D}$. We let the multiset comprised of $N^d$ copies of each request be denoted $\widetilde{R}$. We define the \emph{$r$-degree} of a node $p\in\widetilde{D}$ as follows:
\begin{equation*}
    \deg_{r}(p) \coloneqq \frac{\#\text{ of edges from the $N^{d}$ copies of $r$ connected to $p$}}{\text{total \# of copies of $p$}}.
\end{equation*}
In other terms, it is the ratio of the number of times the online algorithm chooses to match copies of $r$ to $p$, divided by the total number of times $p$ appears in the samples. Next, we define the \emph{$s$-degree}, which for a server $s$, and a given matching (which we will from now on assume to be $\opt_{I}(\tilde{D}, \tilde{S})$), indicates how many times the server $s$ is matched to by the (possibly multiple copies of) node $p$ in the integral matching:
\begin{equation*}
    \deg_{s}(p) \coloneqq \#\text{ of edges from the $N^{d}$ copies of $s_i$ connected to $p$}.
\end{equation*}
Now, let $m(r,s)$ denote the fractional weight from request $r$ to server $s$ (i.e., the fraction of the edge between $r$ and $s$ that is assigned by the algorithm). The overall algorithm, given the $r$ degrees and $s$ degrees generated by the online algorithm $\mathcal{B}$ between $R$ and $\widetilde{D}$ and the precomputed offline optimal matching between $\widetilde{D}$ and $\widetilde{S}$ respectively, matches each request $r$ to server $s$ with the following weight on the edge:
\begin{equation*}
    m(r,s) \coloneqq \frac{1}{N^d}\sum_{p\in\widetilde{D}} \deg_{r}(p)\deg_{s}(p).
\end{equation*}
It then continues this process until all $N$ online requests have arrived, and hence all $N^{d+1}$ corresponding points have been processed by $\mathcal{B}$. It is easy to verify that this algorithm overall matches each request and server in such a way that the sum of all fractional weights of edges incident on them equals one, thus producing a valid fractional matching.
\subsection{Analysis of Algorithm}
We now analyze the performance of the above algorithm. 
\begin{proof}[Proof of~\cref{thm:fractional_matching_guarantees}]
Let $ALG$ denote the above algorithm. Then, in expectation, the cost of $ALG$ is given by:
\begin{align*}
    \cost(ALG)&=\frac{1}{N^d}\left(\beta\cdot \opt_{I}(\tilde{R},\tilde{D})+\opt_{I}(\tilde{D}, \tilde{S})\right)\\
    &\leq \frac{1}{N^d}\left(\beta\cdot \opt_{I}(\tilde{R},\tilde{D})+\opt_{I}(\tilde{R},\tilde{D})+\opt_{I}(\tilde{R},\tilde{S})\right)\\
    &=\frac{1}{N^d}\left(O(\beta)\cdot \opt_{I}(\tilde{R},\tilde{D})+N^{d}\cdot \opt_{I}(R,S)\right)\\
    &=\frac{1}{N^d}\left(N^{d}\cdot O(\beta)\cdot \opt_{F}(R,D)+N^{d}\cdot \opt_{F}(R,S)\right)\\
    &\leq O(\beta)\cdot \opt_{F}(R,D)+\opt(R,S)\\
    &=O(\beta)\cdot \opt_{F}(R,D)+\opt.
\end{align*}
The first line comes from definition of the algorithm, the second line follows from triangle inequality on the second term. The third line comes from the relationship between the ``$N^d$-blowup'' matching (between $\tilde{R}$ and $\tilde{S}$) and the regular matching, and the fourth comes from the integral matching having the same optimal cost as the fractional matching. The next thing to do is to quantify the value of the $O(\beta)$ factor. For online metric matching, the best randomized algorithm against an adversary has competitive ratio $\beta=O(\log^2 k)$, were $k$ is the number of requests to be matched. In this case, we have $k=N^{d+1}$, so $\beta=O(d^2\log^2 N)$. Hence, we have that:
\begin{align*}
    c(ALG)&\leq (d+1)^2\cdot O(\beta)\cdot \opt_{F}(R,D)+\opt_{F}(R,S)\\
    &\leq O(d^2\log^2 N)\cdot \opt_{F}(R,D)+\opt_{F}(R,S).
\end{align*}
However, we are not yet done, as the above quantities involve $D$: but the advice we are given is $\mathcal{A}$, not $D$. Hence, we seek a relationship between $R$ and $\mathcal{A}$. We can do this as follows. First, note that the Wasserstein distance is a factor of $N$ smaller than the cost of the optimal matching. Hence, $\opt_{F}(R,D)=N\cdot W_{1}(R, D)$, where with some abuse of notation, we let the letters denote the set of nodes to be matched on the left, and the distributions induced by them on the right. Then, by the triangle inequality,
\begin{align*}
    N\cdot W_{1}(R, D)&\leq N\cdot\left(W_{1}(R,\mathcal{A})+W_{1}(D,\mathcal{A})\right)
\end{align*}
Recall that $\eta=W_{1}(R,\mathcal{A})$ is the Wasserstein distance between the true requests and the advice distribution error, which is what we wanted to relate our error to. Turning to the remaining term, from the choice of parameters and applying \cref{lem:wasserstein_convergence_rate}, we can bound it as $W_{1}(D,\mathcal{A})=O\left(\frac{1}{N^{1+1/d}}\right)$, from which we get
\begin{align*}
    \cost(ALG)\leq O(d^2 N\log^2 N) \cdot \eta+\opt+o(1),
\end{align*}
where $\opt$ is the cost of the optimal matching between $R$ and $S$. Dividing both sides by $\opt$ gives the robustness claim, and further setting $\eta=0$ gives the consistency claim.
\end{proof}
To extend this result to metric spaces beyond $\mathbb{R}^d$, we can use an embedding result of Linial \etal~\cite{DBLP:conf/stoc/LinialMS98}, combined with an observation by Meyerson \etal~\cite{DBLP:conf/soda/MeyersonNP06} and Bansal \etal~\cite{DBLP:journals/algorithmica/BansalBGN14}.
\begin{lemma}[Linial, Magen, Saks~\cite{DBLP:conf/stoc/LinialMS98}]\label{lem:linial_euclidean_embeddings}
    Let $(\mathcal{X},\dist)$ be a finite metric space on $N$ points, and $D(\mathcal{X},\dist)$ be the least distortion with which $(\mathcal{X},\dist)$ may be embedded in any Euclidean space. Then, $D(\mathcal{X},\dist)\leq O(\log N)$.
\end{lemma}
With this in hand, from any arbitrary metric space, one can induce a metric space defined on a hierarchically separated tree (HST) on $N$ points (where $N$ is the number of servers), paying only a distortion of $O(\log N)$, via the FRT tree embedding method \cite{DBLP:conf/stoc/FakcharoenpholRT03}. This is done by constructing the HST only on the submetric induced by the servers, and when a request arrives, pretend it arrives at the closest server, and matching it online on the tree. By~\cref{lem:linial_euclidean_embeddings}, this metric space on $N$ points can then be embedded into a low dimensional Euclidean space at another distortion cost of $O(\log N)$. This additional total distortion term $O(\log^2 N)$ does not impact the convergence of the $W_{1}(\mathcal{E},\mathcal{A})$ term. Hence, the above algorithm can be adapted for general metric spaces, at the cost of only additional polylogarithmic factors in the $\widetilde{O}(d^2 N)$ coefficient. By this same idea, one can also assume without loss of generality that $d\ll N$: indeed, otherwise one can embed the high-dimensional Euclidean space into a HST, then embed this HST into a low dimensional Euclidean space. 
\subsection{Hardness Results for Randomized Algorithms}\label{sect:hardness_fractional_randomized}
Before proceeding to the advice algorithm for the integral metric matching problem, we finally note that the online fractional metric matching problem is not significantly easier than the integral variant. Namely, we will prove the following theorem:
\begin{theorem}\label{thm:fractional_matching_hardness}
    Any randomized algorithm must have competitive ratio at least $\Omega(\log N)$ for the online fractional metric matching problem against an adversary.
\end{theorem}
Notably, this shows that against an adversary, the online fractional matching problem has the same lower bound as the online integral matching problem.
\begin{proof}
    We define the metric for this problem as the discrete metric. In other words, for two point $i$ and $j$, $\dist(i,j)=1$ if $i\neq j$, and $\dist(i,j)=0$ otherwise. The adversary will behave as follows. The first request $r_1$ will be sent to some point that is distinct from any server. Then, for the remaining requests $r_2,...,r_n$, they will be sent to the server with the highest expected matched capacity at that point in time. Clearly, the optimal solution has cost $\opt(R,S)=1$. To analyze the cost of any randomized algorithm, we can sum the costs incurred at every time step. After the first request comes in, clearly the algorithm must incur a cost of $1$. Then, the adversary will send the second request. By an averaging argument, the server with the highest capacity, $s_2\in S$, has capacity at least $\frac{1}{N}$, and $r_2$ will be sent there. Then, the algorithm will incur a cost of at least $\frac{1}{N}$ matching this request. Similarly, for the third request, by an averaging argument, the server with the highest capacity $s_3$, has capacity at least $\frac{1}{N}+\frac{1}{N(N-1)}=\frac{1}{N-1}$, so the algorithm incurs a cost of at least $\frac{1}{N-1}$ serving this request. Generalizing, on request $i$, the server with the maximum capacity has capacity at least $\frac{1}{N+2-i}$. Hence, the cost of any randomized algorithm is lower bounded by:
    \begin{align*}
        \cost(ALG)&=1+\sum_{i=2}^{N}\frac{1}{N-i+2}
        =O(\log N),
    \end{align*}
    as claimed.
\end{proof}
\section{Online Integral Metric Matching}
In this section, we extend the algorithmic ideas in the fractional matching case to the online integral matching case, proving \cref{thm:integer_metric_matching_algorithm}.
\subsection{Quantization Schemes}
Given the above algorithm for online fractional metric matching, one can attempt to apply the same idea to online integral metric matching. The issue with adapting the above algorithm to the integral case is as follows: At a high level, the above algorithm creates an integral matching on multisets of $N^{d+1}$ points, then divides down into an optimal matching. However, when dividing down the integral matching, there is no guarantee that the resulting matching is integral. Hence, this idea of sampling $N^{d+1}$ points does not work: to ensure an integral matching with this approach, an algorithm can only be permitted to use $N$ points. There is also additional nuance in how one obtains these $N$ points. Naively, one might try to directly sample $N$ points from the advice distribution. However, doing this does not yield an algorithm with good consistency. To see why, consider the case where the advice distribution $\mathcal{A}$ is correct, and the uniform distribution over the requests. If an algorithm simply samples $N$ points from this distribution, with probability (roughly) $1-\frac{1}{e}$, it will miss at least one of the points, and it will not be able to compute the optimal offline matching between requests and servers. Thus one should look for an approach that is more refined than sampling with replacement. One such approach is via a suitable \emph{quantization} of the advice distribution. This approach seeks to approximate the advice distribution $\mathcal{A}$ with a discrete distribution $D$ with bounded support size, in this case $N$. In other words, we can look for a set of points $\alpha$, where $|\alpha|\leq n$, such that if $X$ is a random variable with distribution $\mathcal{A}$, $\alpha$ is chosen such that it minimizes:
\begin{equation*}
    \mathbb{E}\left[\min_{a\in\alpha}|X-a|\right].
\end{equation*}
This set $\alpha$ is commonly referred to as the \emph{N-quantizer}, the \emph{N-codeblock}, or the \emph{N Voronoi partition of order 1} of $\mathcal{A}$ (or simply the Voronoi partition of $\mathcal{A}$), and denoted $V_{N,1}(\mathcal{A})$. We refer to \cite{DBLP:journals/tit/GrayN98} for a survey on quantization, and \cite{10.5555/555805} for the more on the $N$-quantizer. Specifically, the $N$-quantizer obeys the following property:
\begin{lemma}\label{lem:quantizer_wasserstein_distance}
    Let $V_{N,1}(\mathcal{A})$ be the distribution induced by the optimal $N$-quantizer of $\mathcal{A}$. Then:
    \begin{equation*}
        V_{N,1}(\mathcal{A})=\inf_{Q, |\supp(Q)|\leq N}\left\{W_1(\mathcal{A},Q)\right\}.
    \end{equation*}
\end{lemma}
In other words, the distribution induced by the $N$-quantizer will minimize the Wasserstein distance. By \cref{lem:wasserstein_convergence_rate}, we know that the convergence rate of the $N$-quantizer to the advice distribution $\mathcal{A}$ will be upper bounded by $O(N^{1/d})$. However, this approach has benefit of always yielding a consistent algorithm: whilst the $N$-quantizer of a distribution may not be unique, if the advice distribution is a uniform distribution over the $N$ requests, the $N$-quantizer of the distribution will always return the $N$ requests, as opposed to the sampling approach, which misses a point with constant probability. Hence, we will use this quantization approach to generate the $N$ points for the integral algorithm, instead of sampling them from the advice distribution. (As an algorithmic note, finding the optimal $N$-quantizer is equivalent to solving the $N$-centers problem on the advice distribution, and hence any offline algorithm for $N$-centers can be used \cite{10.5555/555805}.)
\subsection{Algorithm for Integral Metric Matching}
We now describe and analyze our algorithm, showing that it achieves the following guarantees:
\begin{theorem}\label{thm:integer_metric_matching_algorithm}
    There is an algorithm for online integral metric matching against an adversary that has cost bounded by $\widetilde{O}(N)\cdot\eta+\opt+\widetilde{O}(N^{1/2})$.
\end{theorem}
The algorithm is the same in spirit as above. The algorithm is given offline the server set $S$, the advice distribution $\mathcal{A}$, and receives the request set $R$ online. Before any online arrivals, the algorithm first computes the  $N$-quantizer of $\mathcal{A}$, and denotes this multiset of points as $D$. It then computes the optimal matching $\opt_I(D,S)$. Then, initialize an online metric matching algorithm $\mathcal{B}$, with competitive ratio $\beta$. As online requests $r_i$ arrive, use $\mathcal{B}$ to match each $r_i$ to some sampled point in $D$. The overall algorithm matches $r_i$ to the corresponding server which was matched to the sampled point in $\opt_I(D,S)$. 

We now analyze the cost of this algorithm, and hence prove \cref{thm:integer_metric_matching_algorithm}.
\begin{proof}
    The cost of the algorithm in expectation is given by:
    \begin{align*}
        \cost(ALG)&=\beta\cdot \opt_I(R,D)+\opt_I(D,S)\\
        &\leq \beta\cdot \opt_I(R,D)+\opt_I(R,D)+\opt_I(R,S)\\
        &=O(\beta)\cdot \opt(R,D)+\opt\\
        &=O(\beta)\cdot N\left( W_1(R,D)+W_1(D,A) \right)+\opt.
    \end{align*}
    The analysis of this algorithm is the same as in the fractional case. However in this instance, since only $N$ points were sampled, $W_1(R,D)$, by \cref{lem:wasserstein_convergence_rate} and \cref{lem:quantizer_wasserstein_distance}, only converges at a rate of $O(N^{-1/d})$. Hence, using the same $\beta=O(\log^2 k)$ as above, we get that for the integral case, the cost of the algorithm is bounded as follows:
    \begin{equation*}
        \cost(ALG)\leq \widetilde{O}(N)\cdot \eta+\opt+\widetilde{O}(N^{1-1/d}).
    \end{equation*}
    In other words, the $W_1(D,A)$ term does not converge fast enough in the integral matching case to vanish, and instead incurs another additive cost. The results in \cref{lem:wasserstein_convergence_rate} are tight up to constants, indicating that with this approach, this additive cost is unavoidable.

    Then, as a final comment, we can use a similar embedding process as in the fractional setting to ultimately embed the matching problem into the low dimensional Euclidean space $\mathbb{R}^2$, which allows without loss of generality the $\widetilde{O}(N^{1-1/d}$ term to be replaced with $\widetilde{O}(N^{1/2})$.
\end{proof}
\begin{remark}
        We briefly comment on the results obtained above with those by Yang \etal~ in \cite{yang2025onlinemetricmatchingworst}. Assuming their setting of advice, where the advice is a set of $N$ points, the above approach does not incur the additive $\widetilde{O}(N^{1-1/d})$ factor, since this setting equates to an empirical distribution over $N$ points, and thus no quantization error from the $W_1(R,D)$ term. The $\widetilde{O}(N)$ coefficient in front of $\eta$ arises as a combination of this being in the adversarial setting compared to the random arrivals setting Yang \etal~ uses, which results in our $\beta$ incurring additional polylog factors, and us using the Wasserstein measure of error, which incurs an additional factor of $N$.
    \end{remark}
\section{Extension to Other Metric Problems}\label{sec:generalization_metric_problems}
Though our use of distributional advice has so far been focused on the metric matching problem, we believe it to be a general framework, of broad applicability. To illustrate this, we discuss in this section how to extend distributional advice to other metric problems, before instantiating it to some concrete examples. Specifically, the types of online optimization problems we are concerned with are tasks which satisfy the following characteristics:
\begin{enumerate}
    \item The problems are online optimization problems where the algorithm knows the value of $N$ (the number of online requests) beforehand, and the cost to be optimized is measured in terms of distance.
    \item While the online requests may be in arbitrary order, the optimal solution does not depend on the order of requests. In other words, permuting the order of online requests should not change the optimal solution.
    \item Given a set of $N$ points denoted $D$ (not necessarily the online requests), the cost of the optimal solution is bounded by the cost of the optimal solution on $D$ plus the cost of the optimal matching from the online requests to $D$.
\end{enumerate}
Some examples which fit the above criteria are problems line online Steiner tree, and online facility allocation: we will elaborate on these in the next subsection. We now describe, at a high level, our framework for adapting distributional advice to such problems.
\paragraph{The Offline Portion.} Fix a distributional prediction $\mathcal{A}$ on a metric space $(\mathcal{X},\dist)$. The algorithm first ``obtains,'' from the advice $\mathcal{A}$, a multiset $D$ consisting of $N$ points. This can be done, e.g., via quantization like above, sampling, or some other appropriate method. Then, the algorithm computes $\mathcal{S}$, the optimal offline solution $\mathcal{S}$ to $D$ as a problem instance.
\paragraph{The Online Portion.} Next, the algorithm processes the online arrivals $r_1,\dots,r_N$. To process them, the algorithm invokes an online adviceless algorithm, and in doing so ``augments'' the optimal offline solution $\mathcal{S}$. What adviceless algorithm is used and what augmenting a solution means will be dependent on the problem. For example, for metric matching, the adviceless algorithm was any online metric matching algorithm, $\mathcal{S}$ was the optimal matching between the advice points and the servers, $\opt(D,S)$, and augmenting $\mathcal{S}$ involved extending this matching to the request set $R$.\bigskip

The first assumption is necessary for the offline phase, as ``obtaining'' $D$ from $\mathcal{A}$ will require knowledge of $N$; the second is implicitly used in the very definition of distributional advice, which does not carry any information about ordering. Finally, the third assumption is key to the analysis, in order to relate the overall cost incurred to the cost of the offline and augmentation steps.

\subsection{Examples of Problems}
Here, we briefly describe two concrete examples online optimization problems satisfying our assumptions, and show how the above framework can be used to adapt distributional advice to them.
\paragraph{Online Steiner Tree.} In the online Steiner Tree problem, $N$ points (the \emph{terminals}) arrive online, and the algorithm must maintain a spanning tree which contains all the $N$ terminals (but may use additional points). The algorithm cannot delete edges which it has already chosen for the spanning tree. The cost of the algorithm is the sum of the lengths of all edges in the spanning tree. Applying the above framework, the algorithm would be given a prediction $\mathcal{A}$ which gives advice about where the $N$ terminals will arrive. From here, using similar techniques, one can obtain $N$ ``advice terminals,'' and compute the minimum spanning tree (MST) on these $N$ advice terminals. This will be the offline solution $\mathcal{S}$. From here, one can use a matching algorithm (e.g., a metric matching algorithm, or one that greedily selects the closest advice point) to augment the solution $\mathcal{S}$. For this problem, it augments the solution $\mathcal{S}$ by adding additional matching edges between the online and advice terminals. We note this approach is very similar to the algorithm given by Xu~\etal~\cite{DBLP:conf/aaai/0002M22}: in their work, instead of assuming distributional predictions, they model the prediction as a set of $N$ advice terminals. Given this advice, their algorithm first computes the offline Steiner tree on the $N$ advice terminals, then augments with additional arrivals.

\paragraph{Online Facility Allocation.} In the online facility allocation problem, the algorithm has a budget (denoted $k$) of online facilities, and $N$ points will arrive in an online fashion. The task for the algorithm is to service the online arrivals by connecting them with facilities. Upon each online arrival, the algorithm can choose to connect them with an already existing facility, or to open a new facility at any location of its choosing (so long as there are fewer than $k$ facilities currently open), and connect the request to the newly opened facility. The allocation of arrivals to facilities, along with the opening of facilities are irrevocable. The cost incurred by the algorithm is the sum of the distances between online arrivals and the facility they were allocated to. Using the above framework, one would start with obtaining $N$ advice points from the advice distribution $\mathcal{A}$. Then, the algorithm would compute the optimal offline solution $\mathcal{S}$, which involves placing $k$ facilities and pairing advice points with facilities so as to minimize the total cost. Then, as the online requests arrive, one can again use a matching algorithm to allocate online arrivals to advice points, and augment the solution $\mathcal{S}$. For this problem, augmenting the solution $\mathcal{S}$ is similar to the metric matching, where the algorithm uses the advice points as intermediaries to determine which facilities the online arrivals should be paired to.
\section{Conclusion and Future Directions}
We extended the study of algorithms with predictions leveraging distributional advice, a type of advice that is arguably more realistic and obtainable than many ad hoc types of predictions previously considered in the literature, and which we believe is a well-motivated and promising research direction. We showed how this type of advice can be leveraged for the online metric matching problem, and gave a framework for how it can be extended generally into metric optimization problems. We list below some concrete avenues for future work:
\paragraph{Removing the Additive Error Term.} Unlike in the fractional matching case, the cost for our integer matching algorithm incurs an additive $\widetilde{O}(N^{1-1/d})$ which arises quantizing a continuous distribution into $N$ points. Whilst this error seems inherent to this approach, this raises the question of whether there could potentially be another way of leveraging the advice which avoids this additive term.

\paragraph{Extending Distributional Advice to Other Problems.} As argued prior, the use of distributions as advice is both novel, yet also natural in the context of obtaining predictions from machine-learning models. It would be of interest to generalize \cref{sec:generalization_metric_problems} further, to broader the class of problems this model can be applied to. Notably, it would also be interesting to see whether a black box theorem similar to the switching lemma in \cite{DBLP:conf/iscopt/AntoniadisBM24} can be developed, %
and to characterize the class of problems to which the approaches in this work apply.
\printbibliography
\appendix
\section{A Hardness Result for Algorithms with Advice}\label{app:hardness_result}
In this section we exhibit a hardness result for both the integral and fractional matching settings.
\begin{theorem}\label{thm:integer_matching_advice_hardness_result}
 Every 1-consistent algorithm with advice must be at least $\Omega(2^N)$-robust against an adversary, in both the fractional and integral settings.   
\end{theorem}
Notably, this is much worse than the traditional bounds for both settings as shown in \cref{tab:compratios}. We first show the result for deterministic algorithms in the integral setting, and the same argument can be used to extend the result to randomized algorithms, and the fractional setting.
\begin{lemma}\label{lem:deterministic_matching_advice_hardness}
    Every 1-consistent deterministic algorithm with advice must be at least $\Omega(2^N)$-robust against an adversary in the integral setting.
\end{lemma}
\begin{proof}
    Consider an instance of the matching problem on $\mathbb{R}^1$, as is illustrated in \cref{fig:integral_matching_hardness}. 
    \begin{figure}[H]
\centering
\begin{tikzpicture}[scale=0.9, transform shape]

\begin{scope}[yshift=3cm]
\foreach \x/\label in {1/s_1,2/s_2,4/s_3,8/s_4}
    \fill[blue] (\x,1) circle (4pt) node[above=3pt] {$\label$};

\foreach \x/\label in {1.5/r_1,2/r_2,4/r_3,8/r_4}
    \fill[red] (\x,0) circle (4pt) node[below=3pt] {$\label$};

\draw[black, line width=0.8pt] (1,1) -- (1.5,0);
\draw[black, line width=0.8pt] (2,1) -- (2,0);
\draw[black, line width=0.8pt] (4,1) -- (4,0);
\draw[black, line width=0.8pt] (8,1) -- (8,0);

\draw[line width=0.8pt] (1,1) -- (2,1) node[midway,above] {1};
\draw[line width=0.8pt] (2,1) -- (4,1) node[midway,above] {2};
\draw[line width=0.8pt] (4,1) -- (8,1) node[midway,above] {4};

\node[left, font=\large] at (0,1) {$R_0$};
\end{scope}

\begin{scope}[yshift=0cm]
\foreach \x/\label in {1/s_1,2/s_2,4/s_3,8/s_4}
    \fill[blue] (\x,1) circle (4pt) node[above=3pt] {$\label$};

\foreach \x/\label in {1/r_4,1.5/r_1,2/r_2,4/r_3}
    \fill[red] (\x,0) circle (4pt) node[below=3pt] {$\label$};

\draw[black, line width=0.8pt] (1,1) -- (1,0);
\draw[black, line width=0.8pt] (2,1) -- (1.5,0);
\draw[black, line width=0.8pt] (4,1) -- (2,0);
\draw[black, line width=0.8pt] (8,1) -- (4,0);

\draw[line width=0.8pt] (1,1) -- (2,1) node[midway,above] {1};
\draw[line width=0.8pt] (2,1) -- (4,1) node[midway,above] {2};
\draw[line width=0.8pt] (4,1) -- (8,1) node[midway,above] {4};

\node[left, font=\large] at (0,1) {$R_1$};
\end{scope}

\end{tikzpicture}
\caption{An illustration of the two possible request sets and matchings for $N=4$. The servers are in blue and the requests are in red, with the subscript of the requests indicating their order. The servers and requests are drawn with different $y$ values for clarity, but in reality the matching occurs in $\mathbb{R}^{1}$.}
\label{fig:integral_matching_hardness}
\end{figure}

For each of the $N$ servers, server $s_i$ will be placed at $2^{i-1}$. In other words, $s_1$ is at $1$, $s_2$ is at $2$, $s_3$ is at $4$, and $s_{N}$ is at $2^{N}$. The adversary will choose one of two request sets, call these sets $R_0$ and $R_1$. The first $N-1$ requests will be the same: $r_1$ will arrive at $\frac{3}{2}$, then $r_i$ will remain at $s_i$ for $i\in[2,N-1]$. Then, in the set $R_0$, the final request $R_N$ will arrive at $s_N$. This results in each $r_i$ being matched with $s_i$, for a cost of $OPT=0.5$. In $R_1$, the final request $r_N$ will instead arrive at $s_1$. This results in the optimal matching for $R_1$ involving matching $r_N$ to $s_1$, and $r_i$ to $s_{i+1}$ for $i\in[1,N-1]$, for $OPT=2^{N-1}-0.5$.
    
    Assume that the algorithm knows the above behavior of the adversary. Namely, it knows both of the possible request sets, and that the adversary will choose one of the two as the real request set. Then, the only relevant advice it can be given is a bit $j\in\{0,1\}$, which is a prediction of which request set it is given, though we make no assumption on what specific form of advice this takes. Then, the algorithm will have no way of determining which request set is the real request set until the final request, and hence it has no way of judging the quality of the advice until the final request arrives. Hence, in order to obtain $1$-consistency, it is forced to trust the advice. Assume that it is given as a prediction the bit $j=1$. Then, upon the arrival of the first request, it will match $r_1$ with $s_2$, and so forth until $r_{N-1}$. Then, assuming this advice is correct, it will correctly match $r_N$ with $s_1$, and this matching will obtain $1$-consistency. However, if the advice is incorrect, the corresponding matching will have a cost of $2^{N-1}-0.5$ against an optimal cost of $OPT=0.5$, and hence result in $O(2^N)$ robustness.
\end{proof}
Clearly, it can be seen that randomization does not help here. Since there is no way to judge the quality of the advice until the last request, a randomized algorithm must still blindly trust the advice for the first $N-1$ requests if it wishes to obtain $1$-consistency, even in expectation, giving the following. 
\begin{lemma}\label{lem:randomized_matching_advice_hardness}
    Every 1-consistent randomized algorithm must be at least $\Omega(2^N)$-robust against an adversary in expectation in the integral setting.
\end{lemma}
Moreover, we now show that allowing the matching to be fractional does not help.
\begin{lemma}\label{lem:fractional_matching_advice_hardness}
    Every 1-consistent algorithm with advice for online fractional matching must be at least $\Omega(2^N)$-robust against an adversary.
\end{lemma}
\begin{proof}
    We use the same advice setting as in the integral case. This means that any fractional matching algorithm still has no way to test the quality of the advice until the arrival of the last request $r_N$. Furthermore, the requirement of being $1$-consistent forbids the algorithm from actually taking advantage of fractional matchings. To see this, consider just requests $r_1$ and $r_2$ in $R_0$. Assume $r_1$ matches a fraction $\alpha$ to $s_1$ and $1-\alpha$ to $s_2$, for a total matching cost of $\frac{1}{2}$. Then, when $r_2$ arrives, it must make a fractional matching that costs at least $\alpha\cdot 1$. The total matching cost at this point is already $\frac{1}{2}+\alpha$, but to be $1$-consistent, the cost must be exactly $\frac{1}{2}$. A similar argument can be done in the $R_1$ case with $r_4$ and $r_1$ on servers $s_1$ and $s_2$. As such, any fractional matching algorithm must make integral matches, if it hopes to be $1$-consistent, and the hardness results from above follow. 
\end{proof}
 Putting together \cref{lem:deterministic_matching_advice_hardness}, \cref{lem:randomized_matching_advice_hardness} and \cref{lem:fractional_matching_advice_hardness}, we prove \cref{thm:integer_matching_advice_hardness_result}. \qed
\end{document}